\documentclass[superscriptaddress,showpacs,preprintnumbers,amsmath,amssymb,psfig,epsfig,12pt]{revtex4}

\usepackage{graphicx}
\usepackage{dcolumn}
\usepackage{bm}
\usepackage{cancel}
\usepackage{slashed}

\def\be{\begin{equation}}
\def\ee{\end{equation}}
\def\bea{\begin{eqnarray}}
\def\eea{\end{eqnarray}}

\begin{document}

\title{ Exclusive Production Ratio of Neutral over Charged Kaon Pair in $e^+e^-$ Annihilation Continuum via `Straton Model'}

\author{Yi Jin}
\affiliation{School of Physics and Technology, University of Jinan, Jinan 250022,
	P. R. China}
\author{Shi-Yuan Li}
\affiliation{School of Physics, Shandong University, Jinan 250100, P. R. China}
\author{Yan-Rui Liu}
\affiliation{School of Physics, Shandong University, Jinan 250100, P. R. China}
\author{Zhao-Xia Meng}
\affiliation{School of Physics and Technology, University of Jinan, Jinan 250022,
	P. R. China}
\author{Zong-Guo Si}
\affiliation{School of Physics, Shandong University, Jinan 250100, P. R. China}
\author{Tao Yao}
\affiliation{School of Physics, Shandong University, Jinan 250100, P. R. China}


\date{\today}

\begin{abstract}
A completely relativistic quark model in the Bethe-Salpter framework 
is employed to calculate the exclusive production ratio of the neutral over charged Kaon pair in $e^+e^-$ annihilation continuum  region for center of mass energies smaller than the $J/\Psi$ mass. The valence quark charge plays the key r\^{o}le. The cancellation of the diagrams for the same charge case (in $K_S + K_L$) and the  non-cancellation of the diagrams for the different charge case (in $K^-+K^+$) lead to the ratio as $(m_s-m_d)^2/M_{Kaon}^2 \sim 1/10$.
\end{abstract}

\maketitle

In low energy region below the mass of $J/\Psi$, it is  generally difficult to employ quark degree of freedom to calculate hadron production, since  perturbative QCD calculation is not valid any more. However, in some simple exclusive processes, we can employ the quarks in the Bethe-Salpter (BS) framework to investigate, provided that the 
 information of strong interaction between quarks  can be absorbed into the BS wave function (BS vertex). As a matter of fact, this idea  emerged just at the time of the birth of quark model,  before  the birth of QCD. One of the most famous examples is the Chinese `Straton Model' in  middle  1960's, namely, `Relativistic Structure Theory Of Mesons And Baryons' \cite{PRINT-67-903, zhu1}.

\begin{figure}[htb]
	\centering
	\begin{tabular}{cccccc}
		\scalebox{0.5}{\includegraphics{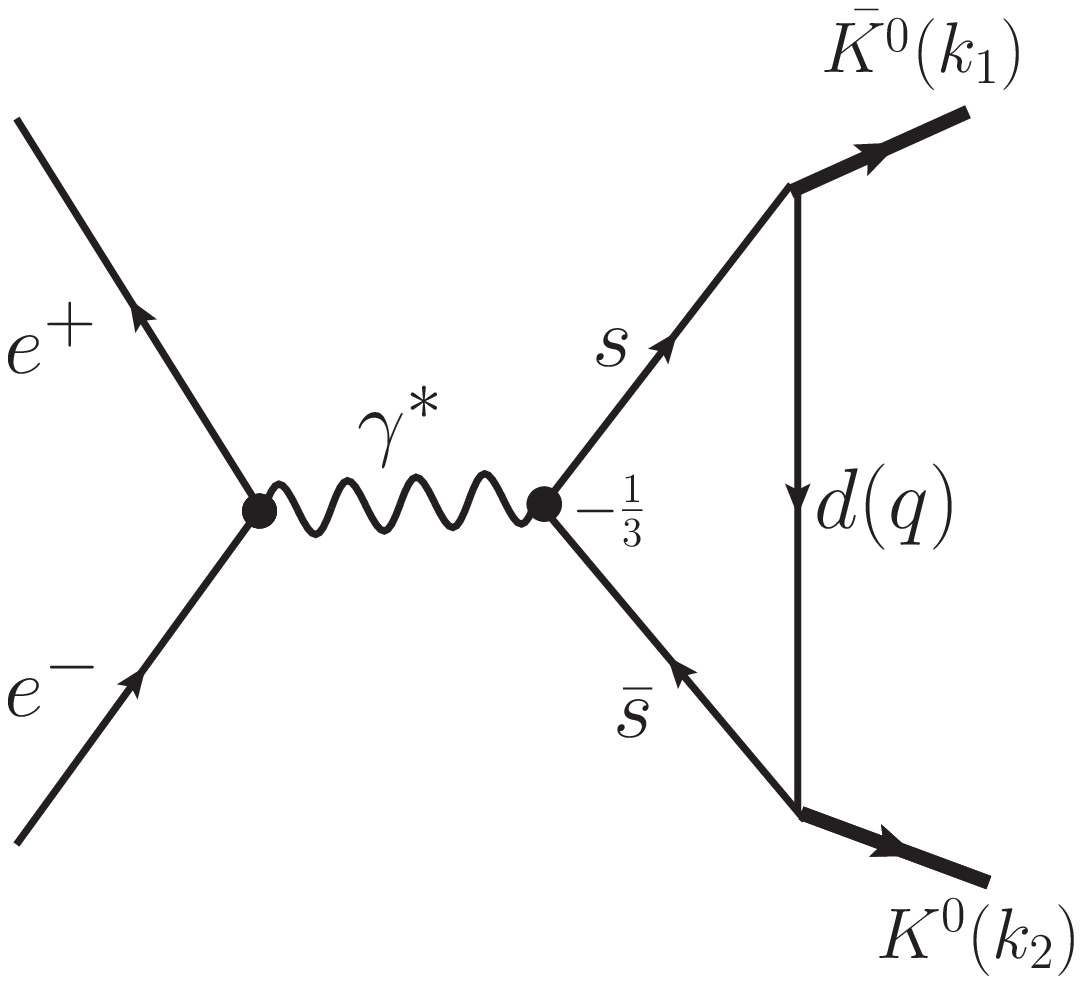}}&
		~~~~~~~~~~&
		\scalebox{0.5}{\includegraphics{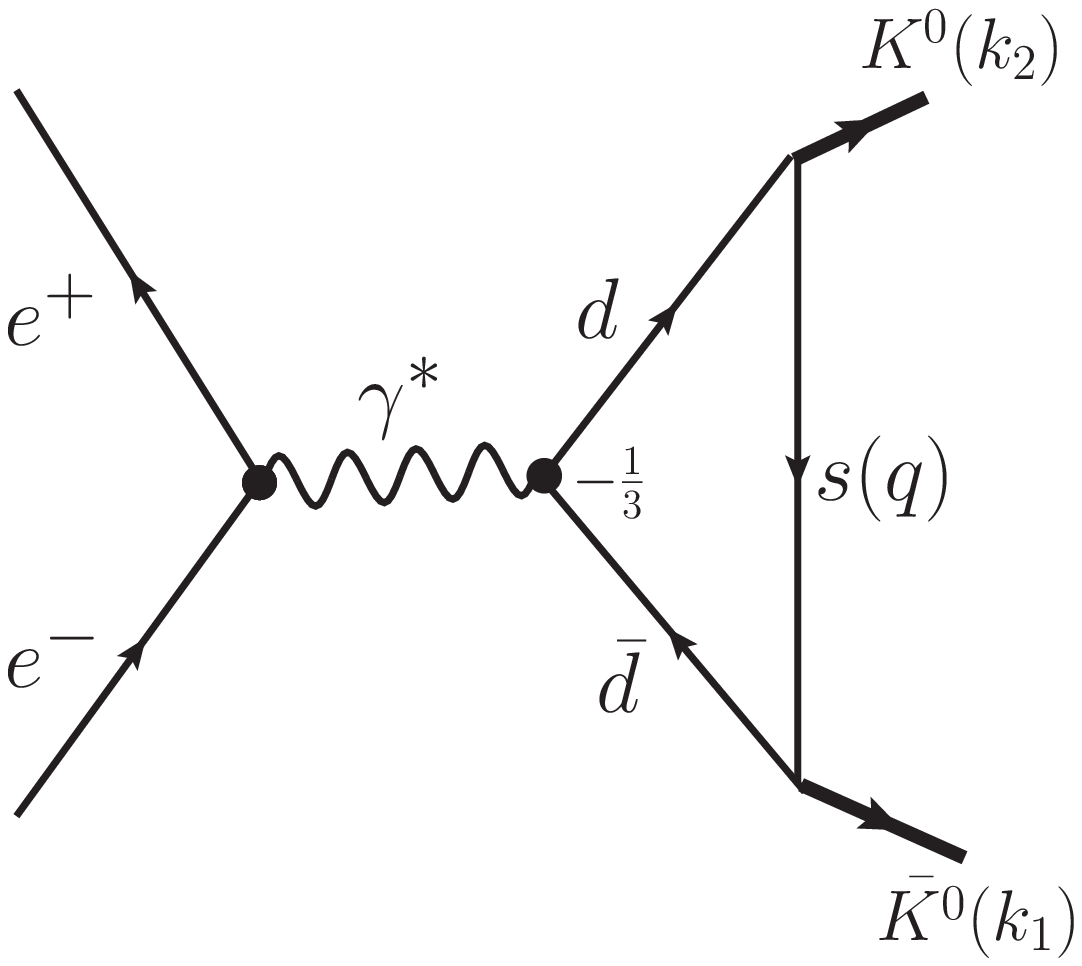}}&\\
		{\scriptsize (a)}&~~~~~~~~~~&{\scriptsize (b)}\\
		\scalebox{0.5}{\includegraphics{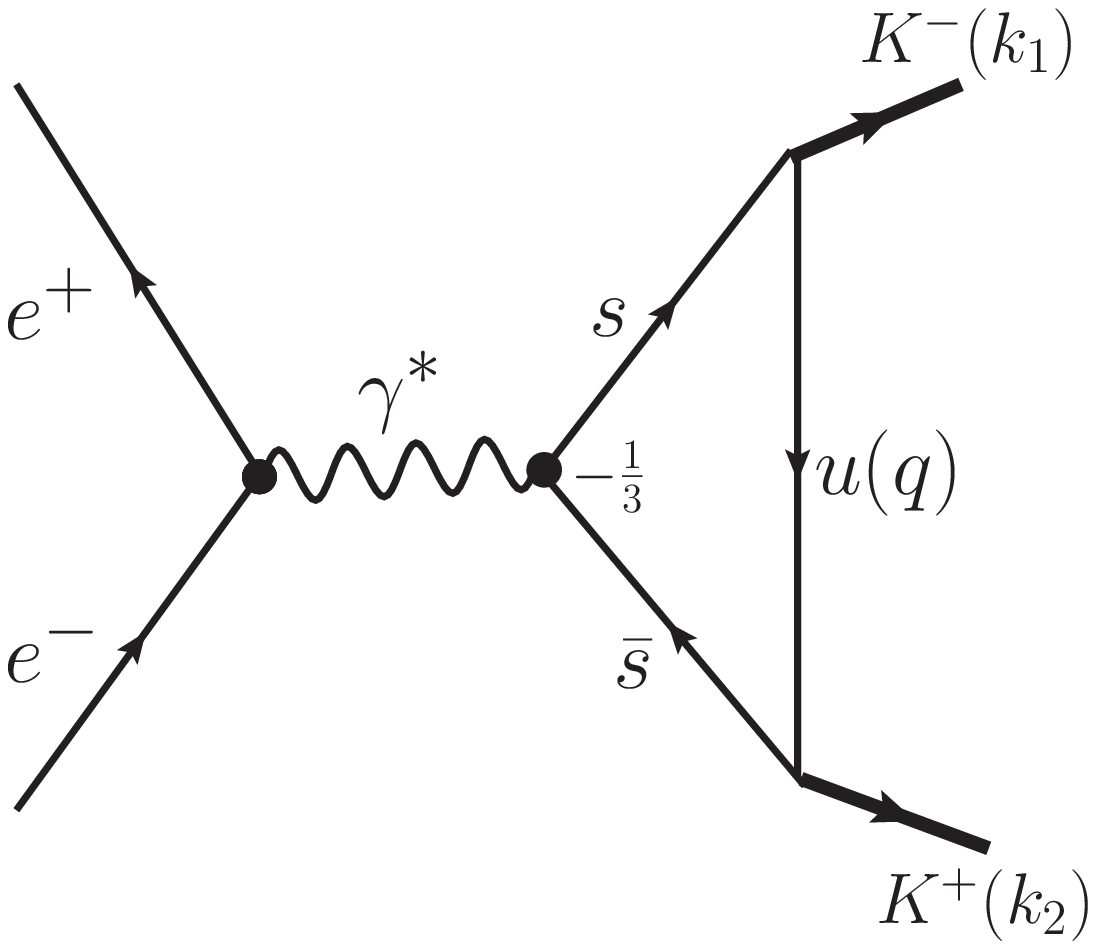}}&
		~~~~~~~~~~&
		\scalebox{0.5}{\includegraphics{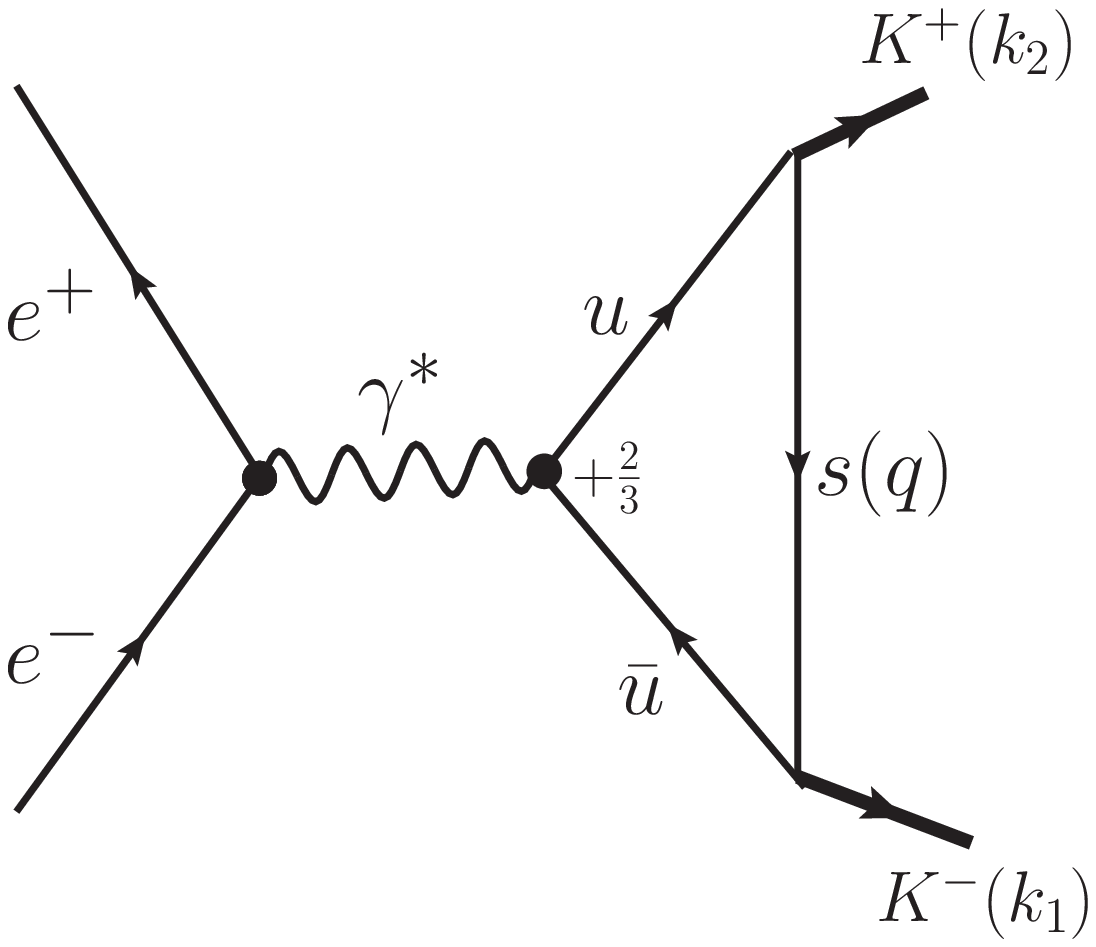}}\\
		{\scriptsize (c)}&~~~~~~~~~~&{\scriptsize (d)}
	\end{tabular}
	\caption{Triangle diagrams for the exclusive productions of $e^+e^- \to K^0 \bar K^0$ and $e^+e^- \to K^+  K^-$. }
	\label{bom}
\end{figure}

 In this paper, we investigate the exclusive production ratio of  the neutral over charged Kaon pair in $e^+e^-$ annihilation continuum for center of mass energies $\sqrt{s}$ smaller than the $J/\Psi$ mass, under the spirit of the Straton Model. More concretely, we employ a completely relativistic BS framework to describe the coupling of the virtual photon to the Kaons via the triangle quark loop (see Fig.~\ref{bom}). The photon-quark vertex is exactly that of the electroweak standard model. The vertices between the quarks and the corresponding Kaon can be given by the BS wave function of the Kaon in terms of `valence' quark field.
 When a resonance is produced in $e^+e^-$ annihilation process, the production ratios of its decay channels at the relevant $\sqrt{s}$ are mainly determined by the properties of the resonance. However, in the continuum region, the mechanism is different.  For our case, the electromagnetic interaction and non-perturbative QCD interaction are
 separately assigned in the diagrams  of Fig.~\ref{bom}.

Taking into account the  scalar function of  the momentum space in the quark-hadron BS vertex,  the loop integral is finite.  So one can calculate the ratio straightforwardly. 
The calculation on the diagrams demonstrates how the naive  suppression on  the production rate of  the neutral Kaon pair  w.r.t. that of the charged ones is obtained. This is just consistent with
the fact that the coupling of neutral particles  with the photon is suppressed  w.r.t. the the case of the charged ones \cite{barbar1}.
The experiment  \cite{barbar1} showes that the suppression can be of one order of magnitude.
 Our calculation obtains the suppression factor as  $ (\Delta m/M)^2$, where $ \Delta m$ is the difference between the masses of the down quark and the strange quark, while $M$ is 
  the mass of the Kaon's.

 %
%



In quantum field theory, the BS wave function $\chi$ is written as
\begin{align}
 \chi(P,q)&=\int \frac{d^4 x}{(2\pi)^4} e^{-iqx}\frac{1}{\sqrt3}\delta_{ij}\langle0|T\psi^i(\frac{x}{2})\bar\psi^j(-\frac{x}{2})|B\rangle
= :S_F(p_1)\Gamma(P,q)S_F(-p_2), \label{1}
\end{align}
and hence the BS vertex $\Gamma(P,q)$ is defined. Here $P$ is the four-momentum of the bound state $|B\rangle$,  while $q$ is the four-momentum to describe the inner movement of the ingredient particles.
They can be expressed as the linear combination by two  four-momenta $p_1$ and $p_2$ of the propagators, $S_F(p_1)$ and $S_F(p_2)$, et vice verse. The only restriction of these momenta is $P^2=M_B^2$, with $M_B$ the mass of the bound state. We emphasize that the ingredient particles can be off mass shell in the BS framework, hence are treated completely as propagators.




This BS wave function can be obtained by fitting corresponding  data for each kind of meson, in term of the valence quark-antiquark field.
To go further, one can set up some certain model for the structure of each kind of multi-states of hadrons, and get the BS wave function by solving the BS equation. In due time some of the model parameters may need to be fixed by data.  Here since we only calculate the ratio, we will not go into the details of the functional of the scalar function $\phi(q^2)$ in the BS vertex, 
\begin{equation}
\Gamma(P,q)=\Omega^{P} \cdot  \phi( q^2).
\end{equation}
Here, the Dirac structure 
\cite{1966,LlewellynSmith:1969az} is arranged as
\begin{equation}
\Omega^{P}=\gamma_{5}-i\gamma_{5}\slashed{P}\frac{B_{1}}{M}
-i\gamma_{5}\slashed{q}\frac{B_{2}}{M}-\gamma_{5}\big {(}
\slashed{P}\slashed{q}-\slashed{q}\slashed{P}\big {)}\frac{B_{3}}{M^{2}}.
\label{eq:2.121}
\end{equation}
In the above equation, $B_{i}$ $(i=1,2,3)$ are the dimensionless coefficients. From this explicit form, a power counting rule  \cite{Bhatnagar:2005vw,Bhatnagar:2006ex,Bhatnagar:2009jg,Bhatnagar:2009mra}  can be directly read out.  
Since we have separated out the factor $\slashed{q}$ from the vertex, the scalar wave function $\phi$ can be taken as a function of $q^2$.
The normalization of the BS wave function is absorbed into $\phi$.


 As in Ref. \cite{moriond}, the amplitude for Fig.~\ref{bom}(a) is written as
\begin{eqnarray}
M_1&=&\frac{-1}{3}\frac{i e^2}{K^2} \overline v_{e+}\gamma_{\mu}u_{e-}
\int\frac{d^4q}{(2\pi)^4} \frac{1 }{\big {[}(q+k_1)^2 -m_{s}^2\big {]}(q^2-m_d^2)\big {[}(q-k_2)^2-m_{s}^2\big {]}}\\ \nonumber
&\times& Tr \big {[} \gamma^{\mu}(\slashed{q}-\slashed{k}_2 +m_{s})\Gamma_2 (\slashed{q}+m_d)\Gamma_1
(\slashed{q}+\slashed{k}_1 +m_{s})\big {]} \phi_{k_1}(q^2)\phi_{k_2}(q^2),
\label{imatrix1}
\end{eqnarray}
where we only consider the leading Dirac structures, i.e. $\Gamma_2 = \gamma^5 (1+ B_1 \slashed{k}_2/M)$ and $\Gamma_1 = \gamma^5 (1+ B_1\slashed{k}_1/M)$ with $K=k_1+k_2$. The other three amplitudes for Fig.~\ref{bom}(b)-(d) are respectively
\begin{eqnarray}
M_2&=&\frac{-1}{3}\frac{i e^2}{K^2} \overline v_{e+}\gamma_{\mu}u_{e-}
\int\frac{d^4q}{(2\pi)^4} \frac{1 }{\big {[}(q-k_1)^2 -m_{d}^2\big {]}(q^2-m_s^2)\big {[}(q+k_2)^2-m_{d}^2\big {]}}\\ \nonumber
&\times& Tr \big {[} \gamma^{\mu}(\slashed{q}-\slashed{k}_1 +m_{d})\Gamma_1 (\slashed{q}+m_s)\Gamma_2
(\slashed{q}+\slashed{k}_2 +m_{d})\big {]} \phi_{k_1}(q^2)\phi_{k_2}(q^2),
\label{imatrix2}
\end{eqnarray}
\begin{eqnarray}
M'_1&=&\frac{-1}{3}\frac{i e^2}{K^2} \overline v_{e+}\gamma_{\mu}u_{e-}
\int\frac{d^4q}{(2\pi)^4} \frac{1 }{\big {[}(q+k_1)^2 -m_{s}^2\big {]}(q^2-m_u^2)\big {[}(q-k_2)^2-m_{s}^2\big {]}}\\ \nonumber
&\times& Tr \big {[} \gamma^{\mu}(\slashed{q}-\slashed{k}_2 +m_{s})\Gamma_2 (\slashed{q}+m_u)\Gamma_1
(\slashed{q}+\slashed{k}_1 +m_{s})\big {]} \phi_{k_1}(q^2)\phi_{k_2}(q^2),
\label{imatrix3}
\end{eqnarray}
\begin{eqnarray}
M'_2&=&\frac{2}{3}\frac{i e^2}{K^2} \overline v_{e+}\gamma_{\mu}u_{e-}
\int\frac{d^4q}{(2\pi)^4} \frac{1 }{\big {[}(q-k_1)^2 -m_{u}^2\big {]}(q^2-m_s^2)\big {[}(q+k_2)^2-m_{u}^2\big {]}}\\ \nonumber
&\times& Tr \big {[} \gamma^{\mu}(\slashed{q}-\slashed{k}_1 +m_{u})\Gamma_1 (\slashed{q}+m_s)\Gamma_2
(\slashed{q}+\slashed{k}_2 +m_{u})\big {]} \phi_{k_1}(q^2)\phi_{k_2}(q^2).
\label{imatrix4}
\end{eqnarray}
Here $(\phi_{k_1}, \phi_{k_2})$ respectively refers to the scalar wave function in the vertex for  $(\bar K^0, K^0)$ or $(K^-, K^+)$, and $m_u, m_d, m_s$ are the masses of the quarks.
In the present study, we adopt the SU(2)  symmetry and thus the scalar wave functions of these four Kaon's are all the same in terms of the inner momentum properly sets. But here we use the loop variable $q$ as self-variable. It is just this choice which makes the functional forms of the $\phi_{k_1}(q^2)$ and $\phi_{k_2}(q^2)$ could be different.

If one only takes the $\gamma ^5$ term in the BS vertex, i.e. taking $B_1=0$, one can find the exact relation $M_1=-M_2$ with a straightforward calculation. So these two diagrams for the $K^0\bar K^0$ case (Fig.~\ref{bom}(a) and (b)) completely cancel. On the contrary, we get $M'_1+M'_2=3 M'_1$ for the $K^+K^-$ case.  Conventionally,  before the power counting rule \cite{Bhatnagar:2005vw,Bhatnagar:2006ex,Bhatnagar:2009jg,Bhatnagar:2009mra} was suggested,  it is considered that the leading Dirac structure is only $\gamma ^5$. If one adopts this, one can conclude that the process  $e^+e^- \to  K_L+K_S$ is vanishing, w.r.t. the process   $ e^+e^- \to K^-+K^+$
 ($K_S$ and $K_L$ are just the perpendicular states constructed as the linear combination  from the other two states  $K^0$ and $\bar{K^0}$).

%

Here we address this  too simplified result, just to call the attention on the fact that, when the coupling to the photon via the valence quark  is the same for the corresponding  two diagrams, they completely cancel for the the vertex taken as $\gamma^5$. This is the case for  the down quark and strange quark having the same charge (Fig. 1 (a) (b)). And because the charge of the up quark and strange quark are different, especially with opposite sign,  the corresponding two diagrams do not cancel, rather are enhanced (Fig. 1 (c) (d)).
However, the power counting rule \cite{Bhatnagar:2005vw,Bhatnagar:2006ex,Bhatnagar:2009jg,Bhatnagar:2009mra} suggests that the leading Dirac structure includes two terms rather than one.
This has  significantly improved the description on the decay constant  
 \cite{Bhatnagar:2005vw,Bhatnagar:2006ex,Bhatnagar:2009jg,Bhatnagar:2009mra}.

Adopting the vertex as $\gamma^5 (1+ B_1 \slashed{P}/M) \phi( q^2)$ (the normalization factor  in $\phi( q^2)$ will be changed), and  by straightforward calculations, one can conclude that   the neutral to charged production ratio is   
\begin{equation}\label{sigratio}
\frac{\sigma(e^+e^-\to K_S K_L)}{\sigma(e^+e^-\to K^+K^-)} \cong (\frac{\Delta m}{M})^2,
\end{equation}
with  $\Delta m=m_s-m_d $.  
The significant contribution by the loop integral on $q$, for $\sqrt{s}$ smaller than the $J/\Psi$ mass, is tamed by the scalar wave function,
leading to its contribution of order of Kaon mass.
In this improved result with the full leading Dirac structure of the vertex, one finds that the cancellation  between  (a) and  (b) in Fig. 1 still works, but not completely, rather, leading to the $\Delta m$ factor.


 %
  In the BS framework, the light quark masses are  parameters.
    Generally the constituent quark mass values are adopted and can work well  \cite{Bhatnagar:2005vw,Bhatnagar:2006ex,Bhatnagar:2009jg,Bhatnagar:2009mra}, hence the value of $\Delta m$ is around 150 MeV.  With the  Kaon mass $M\sim500$ MeV,  we get a rough estimation $(\Delta m/M)^2\sim 1/10$, i.e., one order of magnitude as the experiment \cite{barbar1}   indicated. 
 The above calculations also show that, to get the quantitative result for the ratio between these two exclusive processes,
 especially the  dependence on the difference of the value of the quark masses,
  the full leading Dirac structures  that the power counting rule \cite{Bhatnagar:2005vw} dictates,  is necessary.

    %
 Besides the ground state Kaons,  
 the excited states can also be studied.
 Without referring to the full calculation, but just investigating the relative sign between two contributing diagrams based on  different Dirac structures in the quark-hadron coupling vertex, one can predict the  interesting behaviors: For the case of double vectors, the ratio of neutral over charged is smaller than one, similar as this ground state
 case. But for the case of one pseudo-scalar with one vector, the ratio will turn over to be larger than one. This can be checked by future experiments.




Once precise data is obtained,  the cross section
for the above exclusive processes can be separately fitted,
 to determine the scalar wave function in the vertex, whose explicit form here is not used. Needless to say, a concrete model for the BS wave function to describe
 the  strong interaction in the Kaon's must be introduced. Some more approximation framework is also needed 
 \cite{Bhatnagar:2005vw,Bhatnagar:2006ex,Bhatnagar:2009jg,Bhatnagar:2009mra}  
 for the fitting on the hadron spectroscopy.



We suspect that BESIII can give better experimental measurements in energy
region above $2.00$ ~GeV on the charged and neutral Kaon pair exclusive production. For
example, production cross sections of $K^+K^-$ have been measured at
$\sqrt{s}=2.00 \sim 3.08$~GeV with largely reduced uncertainty comparing to
previous experiments~\cite{Ablikim:2018iyx}. The ratio
$\frac{\sigma{(e^+e^- \to K_S K_L) }}{\sigma{(e^+e^-  \to  K^+K^-)}}$
in our study can be checked
better once the process of $K_S K_L$  exclusive production were soon to be obtained at BESIII, and the scalar wave function fitting both exclusive processes can  also be achieved.



\section*{Acknowledgments}
In writing this paper, we got the sad news that the particle community lost one of the most important founders of the quark model,
M. Gell-Mann. This paper is dedicated for the memorial to him.  We also take this chance to pay our respects to the older generation of Chinese particle physicists.
 This work is supported by National Natural Science Foundation of China (grant Nos. 11635009, 11775130, 11775132, 11605074, U1732263) and the Natural Science Foundation of Shandong Province (grant Nos. ZR2018MA047, ZR2017MA002, ZR2016AM16).



\begin{thebibliography}{99}


\bibitem{PRINT-67-903}


  PRINT-67-903 (by Research Group of the Theory of Elementary Particles, Peking).
  
  \bibitem{zhu1}
  Hung-yuan Tzu,  
Proceedings of the 1980 Guangzhou Conference on Theoretical Physics, 1980, pp. 4-31.

\bibitem{barbar1}
J. P. Lees et al.[BaBar Collab.], Phys. Rev. D 89 (2014) 092002.

\bibitem{1966}
Division of Elementary Particles, Laboratory of Theoretical Physics, Peking University, and 
Laboratory of Theoretical Physics, Institute of Matheamtics, Academia Sinica, 
Acta scientiarum naturalium Universitatis Pekinensi, 1966, 2: 209.

\bibitem{LlewellynSmith:1969az}
  C.~H.~Llewellyn-Smith,
  Annals Phys.\  {\bf 53} (1969) 521.

\bibitem{Bhatnagar:2005vw}
  S.~Bhatnagar and S.~Y.~Li,
  J.\ Phys.\  {\bf 32} (2006) 949.


\bibitem{Bhatnagar:2006ex}
  S.~Bhatnagar and S.~Y.~Li,
  hep-ph/0612084.

\bibitem{Bhatnagar:2009jg}
  S.~Bhatnagar, S.~Y.~Li and J.~Mahecha,
  Int.\ J.\ Mod.\ Phys.\ E {\bf 20}, 1437 (2011).

\bibitem{Bhatnagar:2009mra}
  S.~Bhatnagar, S.~Y.~Li and J.~Mahecha,
  DAE Symp.\ Nucl.\ Phys.\  {\bf 54} (2009) 516.

\bibitem{moriond}
  S.-Y. Li, Proceedings of the 47th Recontres de Moriond (QCD) 2012, P149.




\bibitem{Ablikim:2018iyx}
  M.~Ablikim {\it et al.} [BESIII Collaboration],
  Phys.\ Rev.\ D {\bf 99}, no. 3, 032001 (2019).


\end{thebibliography}
\end{document}